# Upstream penetration of developed counter flow resulting from impingement of multiple jets in crossflow of cylindrical duct

E.V. Kartaev, V.A. Emelkin, M.G. Ktalkherman

Khristianovich Institute of Theoretical and Applied Mechanics

Siberian Branch of Russian Academy of Sciences

It is of special interest in some conventional chemical technologies to provide either fast mixing of reagents or rapid quenching of final chemical product. One of most promising techniques to achieve it is mixing of multiple jets radially injected in crossflow (*JIC*) of cylindrical duct when jets impinge and form a jet flowing towards a crossflow. As result of an interaction of counter-flowing jet and confined crossflow, a recirculation flow zone (*RFZ*) upstream of jets injection plane (*JIP*) is formed (see Fig.1). Of valuable importance is to reveal upstream behavior of developed counter-flowing jet in confined crossflow in order to estimate an extreme *RFZ* size where jets are diluted and mixed under non-equilibrium conditions.

The centerline size $l_p$ of *RFZ* (or penetration depth of counter flow jet) is an axial distance between 2$^{nd}$ stagnation point and *JIP* since 1$^{st}$ stagnation point of impingement of jets is located almost in *JIP* at high momentum-flux ratios $J$ defined as following for variable-density flows:

$$J = \frac{\rho_j V_j^2}{\rho_m U_m^2}, \qquad (1)$$

where $\rho_j$ and $V_j$ are the density and the mean velocity of turbulent jets issued into cylindrical duct, respectively; $\rho_m$ and $U_m$ are corresponding quantities of density and the mass-averaged velocity of crossflow.

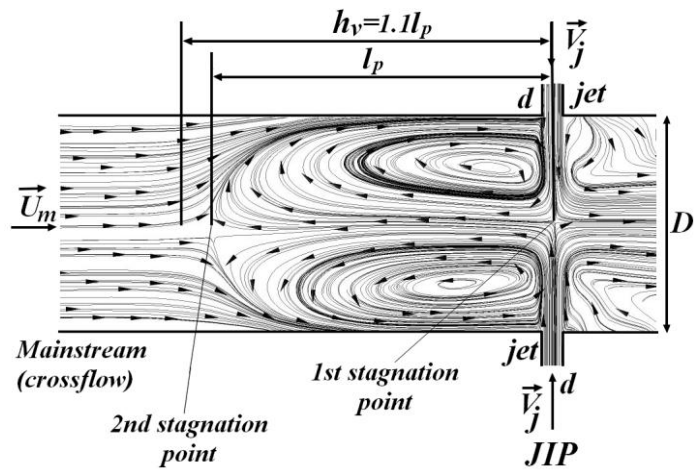

**Fig.1.** Multiple jets-in-crossflow (*JIC*) confined by walls of circular duct.

In our previous studies [1-3] based on experimental data of temperature measurements and numerical simulation a linear dependence between dimensionless parameter of counter flow jet penetration depth $h_v/D$ and square root of $J$ has been obtained:

$$\frac{h_v}{D} \approx K' \frac{d}{DC_d} \sqrt{J}, \qquad (2)$$

where $K'=0.67$, $d$ – orifice diameter, $D$ – cylindrical duct diameter, $C_d=0.8$ – jet discharge coefficient. The term of the counter flow depth of penetration $h_v$ conventionally means the penetration depth of counter flow upstream of *JIP*. It is assumed that counter flow jet reaches a centerline point upstream of *JIP*, when temperature drops minimum to 100 K against undisturbed crossflow temperature at this point. Numerical results indicated that $h_v \approx 1.1\ l_p$.

To describe of the *JIC* mixing quality the parameter of jet radial penetration depth $h/D$ has been proposed in [4,5]:

$$\frac{h}{D} = K \frac{d}{DC_d} \sqrt{J}, \qquad (3)$$

here $K=1.7$. This value does not make any physical sense at $h/D > 0.5$ when radially injected jets impinge. Nevertheless, it is assumed that $h/D$ would govern the intensity of impinging jets interaction and hence influence the mixing quality and rate. In [6,7] it was demonstrated that, as the geometrical similarity of the mixing chambers, working in the impinging jets mode, being followed, the mixing quality is almost constant at the same values of the penetration parameter $h/D$. Thus, the parameters of the counter flow axial penetration depth $h_v/D$ and radial penetration depth $h/D$ are related as follows from (2) and (3) at $J < 625$ [2,3]:

$$h_v/D \approx (0.33 \div 0.44) h/D. \qquad (4)$$

*Experimental setup*

This study has been carried out based on experimental setup and operating modes described in [2,3] (see Fig.2). The heated nitrogen was used as crossflow gas of the cylindrical duct of diameter $D=32$ mm. Room-temperature air (or nitrogen) was used as an injected jet fluid. Cooling gas was fed tangentially into the annular manifold, its direction is shown in Fig.2 by arrows. Then it is injected through $n=8$ holes of $d=3$ mm in diameter perpendicularly to the mainstream. Upstream of *JIP*, 4 thermocouples (type K, chromel-alumel) were set to measure the temperature on the duct centerline. Each thermocouple was turned about the previous one to the angle of 90º in order to decrease the disturbance imposed by them in the flow. For one of implemented configuration of type K positioning with inserted supplementary section the centerline distances between the junctions of thermocouples and *JIP* were -62 (or -56), -75, -88, and -101 mm, respectively (see Fig.2), here $x<0$ upstream of *JIP*.

In contrast to [2,3], orifices of 3 mm in diameter instead of 5 mm were chosen in order to reach greater values of *J* at the same jets' flow rates. In turn, this should allow us to indicate gradual deceleration of counter flow jet and finally its complete stop at very large values of *J*.

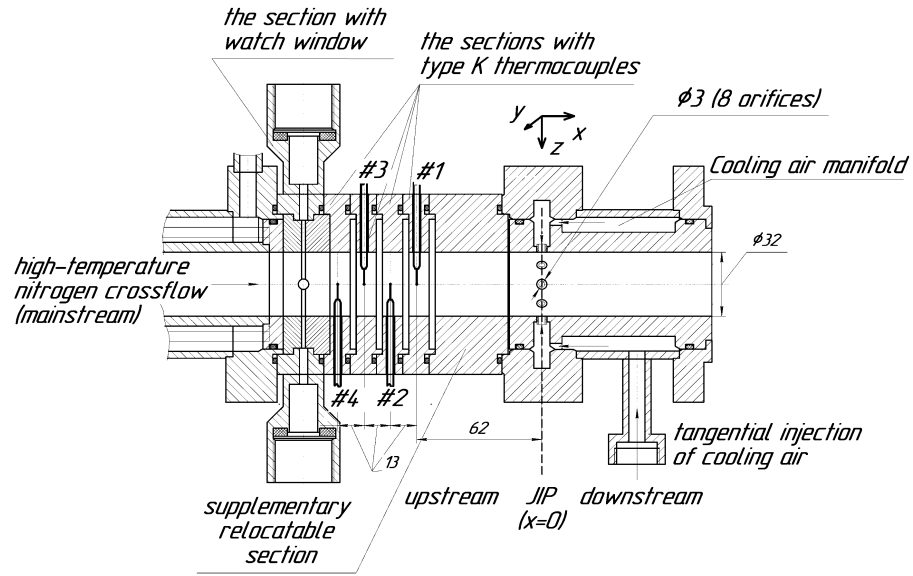

**Fig.2.** Schematic of mixing section of cylindrical duct with one of implemented configuration of type K positioning and the section of annular manifold as well as the supplementary section (*d/D/n* = 3/32/8). All sizes in mm.

Reducing the orifice diameter resulted in a necessity of more precise "focusing" of turbulent jets issuing into the cylindrical duct. It has been provided by increased orifice wall thickness *t* (*t/d*=1.33 or 1.67 compared with *t/d*=0.60 in [2,3]) and similar flow pattern at the entry of each orifice.

*Results*

Fig.3 shows typical temperature profiles upstream of *JIP* obtained at various *J* (various flow rates of cooling jets) at preset configuration of type K thermocouples positioning shown in Fig.1.

Since *x* axis is co-directional with mass-averaged crossflow velocity $U_m$ as depicted in Fig.2, centerline coordinates *x/D* < 0 upstream of *JIP*. It is seen as *J* grows as the upstream temperature decreases. It means that counter flow jet penetrates deeper towards the crossflow along cylindrical duct centerline. It is also clear that there is an asymptotic value of $h_v/D$. This follows from comparing curves 2 and 3, since valuable increase of *J* does not lead to deeper penetration of counter flow jet – temperature at the centerline location of thermocouple #3 (*x* = -88 mm) is almost the same.

In Fig.4 the experimental data concerning $h_v/D$ as a function of square root of *J* obtained in this work (*d/D/n* = 3/32/8) as well as in our previous studies [2,3] (*d/D/n* = 5/32/8) are depicted with corresponding fitted dotted lines.

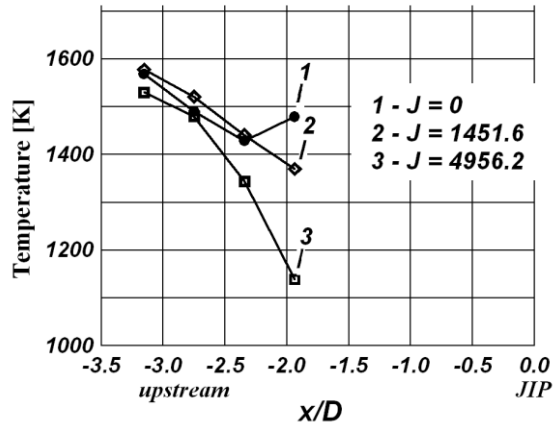

**Fig.3.** Cylidrical duct centerline temperature profiles as a function of non-dimensional distance $x/D$ upstream of *JIP* ($x/D = 0$) at various $J$ at given configuration of type K thermocouples positioning.

The latter demonstrates only linear dependence $\sim J^{1/2}$ according to formula (2). The former turned out to consist of linear region $\sim J^{1/2}$ (or $\sim h/D$) within $4 < J^{1/2} < 30$ (formula (2)), the non-linear region $\sim J^{1/6}$ (or $\sim (h/D)^{1/3}$) within $30 < J^{1/2} < 60$, and asymptotic one $h_v/D \approx 2.3$ at $J^{1/2} > 60$. So, it can be emphasized that for given mixer geometry centerline size of *RFZ* cannot exceed 2.3 of duct diameter.

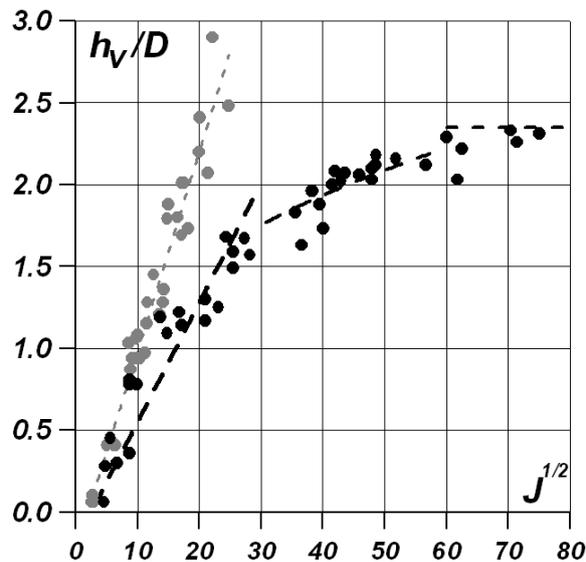

**Fig.4.** The parameter of counter flow jet penetration depth $h_v/D$ as function of square root of $J$: black circles – experimental data of present tests ($d/D/n = 3/32/8$); separate black dotted lines correspond to linear region ($4 < J^{1/2} < 30$), non-linear region ($30 < J^{1/2} < 60$) and asymptotic region ($J^{1/2} > 60$). Gray circles – experimental data obtained in [2,3] ($d/D/n = 5/32/8$).; gray dotted line fitted to experimental points [2,3].

The obtained complicated dependence $h_v/D = f(J^{1/2})$ looks very similar to that discovered in [8,9] when counter-flowing jet issues from centerline-located tube toward a mainstream flow of cylindrical duct (equi-density flows).

*Conclusion*

The results of present *JIC* experimental study for given mixer geometry indicates that parameter of centerline counter flow penetration depth $h_v/D$ as a function of square root of $J$ consists of three distinct parts: linear region $\sim J^{1/2}$ (or $\sim h/D$) within $4 < J^{1/2} < 30$, the non-linear region $\sim J^{1/6}$ (or $\sim (h/D)^{1/3}$) within $30 < J^{1/2} < 60$, and asymptotic one $h_v/D \approx 2.3$ at $J^{1/2} > 60$.